# *Element- and enantiomer-selective visualization of ibuprofen dimer vibrations*

R. Mincigrucci[1,*,⊥], J. R. Rouxel[2,#,⊥], B. Rossi[1,3], E. Principi[1], C. Bottari[1,4], S. Catalini[5], D. Fainozzi[1,4], L. Foglia[1], A. Simoncig[1], A. Matruglio[6], G. Kurdi[1], F. Capotondi[1], E. Pedersoli[1], A. Perucchi[1], F. Piccirilli[1], A. Gessini[1], M. Giarola[7], G. Mariotto[8], S. Mukamel[9], F. Bencivenga[1], M. Chergui[2,^], C. Masciovecchio[1,+]

⊥ Authors contributed equally

1 Elettra Sincrotrone Trieste S.C.p.A., Strada Statale 14 - km 163,5 in AREA Science Park 34149 Basovizza, Trieste ITALY

2 Laboratoire de Spectroscopie Ultrarapide (LSU) and Lausanne Centre for Ultrafast Spectroscopy (LACUS), École Polytechnique Fédérale de Lausanne, CH-1015 Lausanne, Switzerland

3 Department of Physics, University of Trento, Via Sommarive 14, 38123, Povo, Trento

4 Department of Physics, University of Trieste, Trieste, Italy

5 European Laboratory for Non-Linear Spectroscopy (LENS), Università di Firenze, 50121 Florence, Italy

6 CERIC-ERIC Strada Statale 14 - km 163,5 in AREA Science Park 34149 Basovizza, Trieste ITALY

7 Centro Piattaforme Tecnologiche, University of Verona, Policlinico GB Rossi, Ple. L.A. Scuro, 10, 37134 Verona (Italy)

8 Dipartimento di Informatica, Università di Verona, Strada delle Grazie 15, 37134 Verona (Italy)

9 Department of Chemistry and physics and astronomy, University of California Irvine, Irvine, California 92697, United States

\* riccardo.mincigrucci@elettra.eu

\# jrouxel@uci.edu

^ majed.chergui@epfl.ch

\+ claudio.masciovecchio@elettra.eu

**In chemistry, biology and materials science, the ability to access interatomic interactions and their dynamical evolution has become possible with the advent of femtosecond lasers[1]. In particular, the observation of vibrational wave packets via optical (UV-visible-IR) spectroscopies has been a major achievement as it can track the motion of nuclei within the system[2–4]. However, optical spectroscopies only detect the effect of the interatomic vibrations on the global electronic surfaces. New tuneable, pulsed and polarized sources of short-wavelength radiation, such as X-Ray Free electron lasers, can overcome this limitation, allowing for chemical and, of primary importance in biochemistry, enantiomeric selectivity. This selectivity may be complemented by taking into account the chemical shifts of atoms belonging to different molecular moieties.**

In biochemical reactivity, such as protein-target (enzymes, RNA, micro RNA) interactions the low-frequency regime (< 300 cm$^{-1}$) of the vibrational spectrum is mainly responsible for and the most affected by the biological function, as it reflects the global and continuous changes in the molecular geometry[5]. Furthermore, there is a need to detect and map the motion of light atoms, such as C, N, O, involved in molecular vibrations with periods as slow as a few 10's of ps (i.e. vibrational frequencies down to below few cm$^{-1}$), as this allows to address various binding processes and, thereby, achieve site-specificity. Steady-state and ultrafast vibrational terahertz (THz) to infrared (IR) spectroscopies, are the most common methods used to access ground state vibrational modes, but they are limited to dipole-allowed vibrational transitions, while disentangling near-degenerate vibrational modes can be challenging. Steady-state Raman spectroscopy is characterised by weaker signals and it also obeys selection rule, in addition to the requirement of an efficient suppression of the elastic peak at low frequencies[6,7]. One approach for accessing low frequency modes in the time-domain is impulsive stimulated Raman scattering (ISRS)[8]. It exploits the fact that the very short pump pulse is spectrally broad, encompassing several vibrational levels of the ground electronic state and therefore stimulating Raman transitions to the ground state, which thereby create a coherent superposition of vibrational states, i.e. a ground state wave packet. Monitoring the wave packet dynamics in the time domain can circumvent the limitations caused by the finite spectral resolution of steady-state spectroscopies. Low frequency modes then become easier to detect because of their long oscillation periods. While this is of great value for getting insights into, e.g., the structure of protein/nucleic acid complexes, optical radiation is not element-specific. Adding atomic level specificity would bring the advantage of identifying which elements are involved in the conformational rearrangements during the formation of complexes and to unravel dynamical aspects of the interactions between residues and bases involved in bio-recognition.

This is possible using short-wavelength (hard X-rays to extreme ultraviolet) radiation that can be tuned to specific core-transitions of the various atomic elements. In recent years short-wavelength spectroscopies have successfully been extended into the ultrafast regime[9], allowing the observation of vibrational wave packets around specific atoms in molecular systems[10]. Femtosecond X-ray absorption or emission spectroscopy can map the time evolution of interatomic vibrations with element-selectivity but this approach is limited to systems containing heavy atoms that have relatively low frequency modes, and efficiently absorb or scatter hard X-rays.[10,11]

Another important aspect for biological systems is enantiomeric selectivity. Most biological molecules are chiral, i.e. they exist in two different forms, called enantiomers, that have the same chemical composition but are mirror images of each other. However, biological activity is generally homo-chiral and therefore distinguishing between enantiomers is a central issue in pharmacology, toxicology and medicine. The method most commonly used to detect enantiomers is circular dichroism (CD) spectroscopy, which exploits the fact that light polarized into a circular wave is absorbed differently by left-handed and right-handed enantiomers. In the spirit of monitoring the evolution of biological systems, sub-picosecond to nanosecond time-resolved CD optical spectroscopies have been implemented in various spectral regions,[12–14] in order to study the absorption bands of amino-acid residues, nucleobases and peptide chains[14,15]. Extending these capabilities to core-level spectroscopies allows to discriminate biological activities (binding, reactivity) with atomic-specificity within a selected enantiomer. Simulations have shown that X-ray CD signals of molecular systems vary with the electronic coupling to substitution groups, the distance between the X-ray absorbing element and the chiral centre, as well as geometry and chemical structure[16]. Combining CD spectroscopy with element-selectivity of light atoms such as C, N, O is particularly attractive for the study of biological systems.[17] Since the core transitions of these elements lie between 280 eV (C) and 530 eV (O), this calls for ultrashort sources of circularly polarized soft X-ray pulses. Table-top sources based on High Harmonic Generation are not routinely implemented in this photon energy range, and the control of their polarization has still to be demonstrated. [18,19] On the other hand, circularly polarized soft X-ray pulses can routinely be generated at the FERMI free electron laser in the region of the carbon K-edge[20].

Here we demonstrate an element-specific ultrafast soft X-ray absorption experiment that allows to visualize and disentangle low-frequency nearly degenerate vibrational modes involving specific Carbon atoms in a racemic mixture of Ibuprofen (IBP, 4-isobutyl-2-phenylpropionic acid). We furthermore demonstrate how polarization control of the EUV pulses adds enantiomeric selectivity, tying it together with the element-selectivity and mode-specificity.

IBP is an over-the-counter anti-inflammatory non-steroidal drug that is widely used.[21] In the solid racemic mixture, the two enantiomers, labelled (S+)- and (R-), form a cyclic dimer through intermolecular hydrogen bonds between the carboxyl groups of two adjacent molecules[22], as shown in Figure 1a. The conformational stability of this dimeric arrangement has been revealed by X-ray diffraction and Raman scattering.[23] IBP was first used as a racemic mixture but later, focus of the pharmaceutical industries shifted to the (S+)-ibuprofen when it was found that this form enhances the effect of analgesia in animals (including humans), more rapidly than the (R-)-enantiomer does. The bio-activity of the two enantiomers, and in particular the cross-monomer allosteric inhibition, in which S-IBP can competitively block the action of one monomer of the cyclo-oxygenase (COX) enzyme, composed by two equal halves, is however not fully understood.[24]

Recent low-frequency Raman studies[25] have attributed spectral features around 25 cm$^{-1}$ to intermolecular vibrations. We repeated these measurements (see § S5.2 for details) and reproduced the results, while a weak additional band appears at ~28 cm$^{-1}$ (see Figure S8). We further recorded Fourier Transform Infrared (FTIR) spectra of the samples (see § S5.2 for details), which also exhibit the presence of a feature between 20 and 30 cm$^{-1}$ (see Figure S9). To interpret the Raman spectrum, we used Density Functional theory (DFT) calculations described in § S4.1. These reveal that the lowest frequency Raman band consists of three near-degenerate intermolecular vibrational modes (Table 1), which characterise different conformations of the global dimeric structure as depicted in Figure 1 (an animation is given in the SI). The lowest frequency mode ($\omega_1$ = 21.5 cm$^{-1}$), is more localized on the S-dimer and involves a rotation of the benzene ring around the 2-5 axis, i.e. a torsional deformation of both the phenyl ring and the isobutyl group (Figure 1b). The $\omega_2$ = 22.6 cm$^{-1}$ mode involves an out-of-plane twisting (the plane defined by the carboxylic groups) accompanied by a C=CH$_3$ stretch (Figure 1c). Similar to the lowest frequency mode, in the R-dimer (Figure 1d), the mode at $\omega_3$ = 28.8 cm$^{-1}$ involves large displacements of the same molecular groups, in the form of a rotation of the benzene ring around the 18-21 axis and the close isobutyl group. Low-frequency vibrational wave packets, made out by the three low-lying $\omega_1$ to $\omega_3$ modes, can be generated by ISRS on crystallized ibuprofen and are probed here by carbon K-edge absorption spectroscopy.

The experimental scheme used in the present study is shown in Figure 2. An ultrashort pulse at 4.7 eV (263.8 nm) with temporal and spectral full widths at half maximum (FWHM) of ~80 fs and ~40 meV, respectively, excites the system resonantly in order to enhance the cross section of the ISRS process (see §S.2 for details). A soft X-ray probe pulse (~30 fs duration and ~250 meV FWHM) is then used to monitor the changes in transmission across the carbon K-edge absorption. A reference steady-state carbon K-edge spectrum of the system was recorded for left (LC) and right circularly (RC) polarized light on a pure S-IBP sample. The procedure to scan point by point across the C K-edge spectrum with the FEL is described in § S2. The spectra are shown in Figure 3a and they exhibit clear differences. In particular, the spectrum obtained using LC light shows an enhanced absorption at the edge (285-286 eV) compared to the RC counterpart, while above 287 eV, the RC light has a higher absorption. The difference between these two spectra is shown in Figure 3b, and it represents the Carbon K-edge circular dichroism of the system.

In order to better grasp the contribution of each carbon atom in the K-edge spectra, we have simulated them using quantum chemical calculations at the cc-pVDZ/RASSCF(9/8) level (details are given in § S4). The calculations were performed for four different active spaces (AS's) until convergence was reached (see § S4.3). Table S2 gives the K-edge energies of the different carbon atoms as numbered in Figure 1a, while Figures S6 and S7 show their energies and oscillator strengths in the form of stick diagrams for the different AS's. AS3 is taken as the most reliable AS because it shows results comparable to AS2 but is larger than the latter. A rigid shift of 6.4 eV was applied to the core transition energies in order to match the experimental spectra. The elements that come close to the edge energy around 285 eV are atoms (17,3) and (23, 7), (24, 8), (25, 9), (28, 12) and (29, 13). This labelling refers to the identical atoms in the two enantiomers. The energy difference between the different C atoms is ascribed to the K-edge chemical shift due to different local environments around the atoms (see § S4.3 for details). However, of these only (28,12) near ~285.7 eV and (17,3) near ~285 eV have an appreciable oscillator strength (Figure S7). In Figure 3a, we reproduce the stick diagram due to these atoms. Therefore, with energy-resolution we can selectively single out specific atoms in the system, which have appreciable oscillator strength. Adding the probe circular polarization further adds the selectivity to the enantiomer to which the atom belongs. Therefore, the combined use of photon energy and polarization extends the site-selectivity by permitting to access a specific atom of the molecule.

In the present case, the selected atoms belong to the regions of the molecule most affected by the highlighted dimer vibrations (see Figures 1 b-d and animation in the SI), which efficiently modulate the absorption of the investigated atoms, in the time-resolved signal[10,26].

Figure 4 shows the time evolution of the signal upon impulsive excitation of the system at ~4.7 eV and for different soft X-ray probe photon energies and polarizations. All traces reveal periodic intensity modulations (5 - 10 % amplitude of the transmission signal) around a mean value of the transmission. The data in panel a) were obtained at a probe energy of 285.7 eV and with RC polarized light. Panels b) and c) show time scans acquired using a probe photon energy of 285.0 eV with LC and RC polarization, respectively. Panels a) and b) exhibit a clear modulation with a sine dependence, as expected for Raman-induced processes[27], and a ~1.5 ps period. Despite the lower signal-to-noise (S/N) ratio, a periodic modulation can still be distinguished in panel (c), but it is quickly damped. The lower S/N in the latter case is due to the fact that the CR absorption signal at 285 eV is weak and it rides on a background of increasing continuum absorption, which extends towards higher photon energies (see Figure 3a). This should be contrasted with the CL signal at the same energy. The three traces were fitted to damped sinusoidal functions (described in § S3), which yield the frequencies and damping constants given in tables 1 and S3. These frequencies are in agreement with both the calculated and measured lowest Raman peaks, despite the large error bars. The damping times of these three modes correspond to 10 to 20 $cm^{-1}$ spectral widths, which explains why they could not be resolved in the steady-state Raman spectrum, where a high instrumental resolution may be washed out by the intrinsic line width. See e.g. § S5.1.

It is remarkable that despite the ability to generate and a priori observe wave packets of modes up to ~300 $cm^{-1}$ with our pump-probe cross-correlation, the experiment seems selective to only low-frequency modes. Indeed, modes up to 300 $cm^{-1}$ (and beyond) have been reported by non-resonant Raman spectroscopy[25,28]. In resonance Raman scattering, the modes that show up on the ground state surface are the templates of those generated in the excited one. If the latter undergoes few selective deformations, only these will show up in the spectrum[29]. This scenario is confirmed by a recent calculation of the resonance Raman spectrum of the IBP monomer under 4.663 eV excitation[30], which shows signals in the <50 $cm^{-1}$ range, while no further resonance Raman activity appears before 400 $cm^{-1}$. Thus, the choice of the pump energy determines, which modes are excited. These happen to be the low-frequency ones, which most affect the global deformations of the dimer. In our experiment, the combination of element- and enantiomeric-selectivity identifies which atoms are most responsive to these modes, and the oscillations seen in Figure 4 are likely due to both an energy and an intensity modulation of the selected C-atoms. Indeed, the molecular orbital (MO) character of the core levels transitions discussed in § S4.3, implies that they are prone to geometry changes of the molecule which affects the value of the transition energies and oscillator strengths. This is also more the case with low frequency vibrations, which modify the entire molecular edifice, than with high frequency ones that generally have small amplitudes and are more localised.

The present result suggests a strategy to observe the atoms that are most affected by vibrational modes in a given enantiomer and call for more systematic investigations. This also pave the way for the direct investigation of the drug/target intermolecular vibrational dynamics, with the potential to understand the marked differences in biological activities of enantiomers, or easily follow the dynamic of metal complexes in medical applications[31]. More generally, such detailed level of understanding may lead to new design strategies of bioactive molecules such as vibrational/deformational engineering by the use of isotopes to modify the vibrational behaviour of an identified atom or molecular moiety but keeping unaltered the electronic properties. Finally, an actinic pulse could be added to the present super-selective pump-probe approach, triggering photoactive reactions and tracking the behaviour of selected atoms during the chemical process.

Acknowledgments: MC acknowledges support by the Swiss NSF via the NCCR:MUST and by the European Research Council Advanced H2020 Grant ERCEA 695197 DYNAMOX. S.M gratefully acknowledges the support of National Science Foundation (CHE-1663822) and of the Department of Energy (DE-FG02-04ER15571).

Authors' contributions: RM, EP, CB, SC, LF, AS, GK, FC, EP, AG, FB conducted the FEL experiment; JRR performed core state calculations; BR performed Raman simulations; RM and DF analyzed the experimental data; RM, JRR, BR, DF, SM, MC, CM discussed and interpreted the data; AM and RM prepared the substrate hydrophilization and deposited the sample; AP and FP measured the IR spectrum of the IBP powder; BR, MG, GM measured the low frequency Raman spectrum of the IBP powder; RM, JRR, BR, MC, CM wrote the manuscript; all the authors revised the manuscript.

**Table 1**: parameters of the fit of the temporal traces using damped sine functions (see § S3 for details). The low frequency steady state Raman spectrum exhibits only one band centered around 25 cm$^{-1}$ whose width increase from ~5 cm$^{-1}$ at 100 K to ~10 cm$^{-1}$ at 300 K [25].

| Probe energy/polarization | Frequency (cm$^{-1}$) | Damping constant (ps) | Calculated (cm$^{-1}$) |
|---|---|---|---|
| 285 eV  CL | 23.9 ± 1.1 | 1.7± 1 | 21.5 |
| 285 eV CR | 29 ± 0.7 | 2.8 | 28.8 |
| 285.7 eV CR | 24.3 ± 0.9 | 2.5 ± 2 | 22.6 |

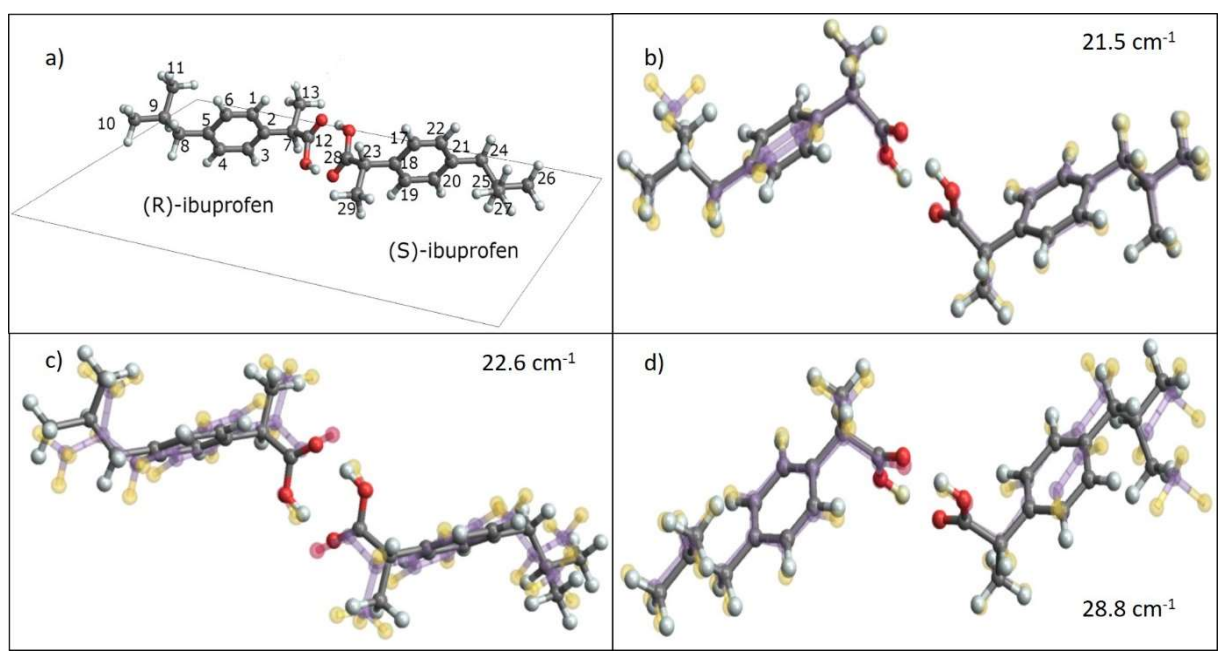

**Figure 1 –a)** The Ibuprofen dimer with the numbering of carbon atom. Panels b to c show the three intermolecular modes of the dimer (see animation in the SI). The structures appearing in contrasted and semi-transparent colours, correspond to the minimum and maximum deformations for the calculated intermolecular modes of the dimer.

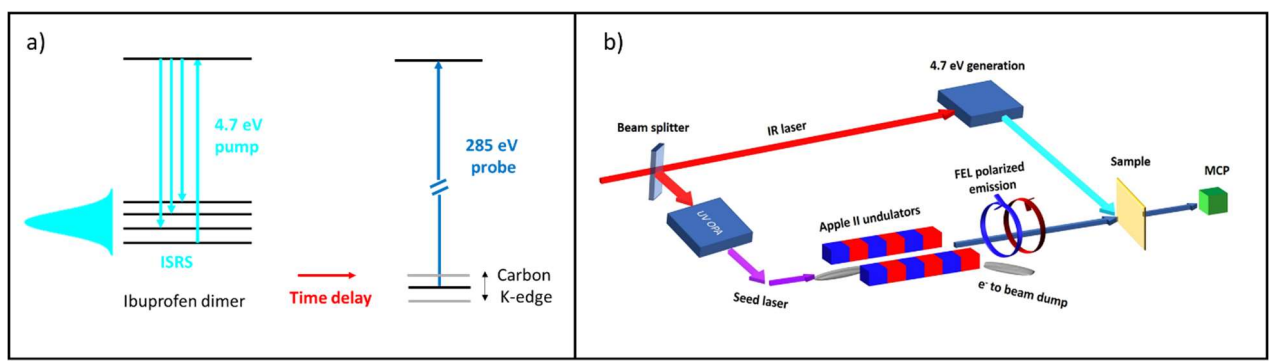

**Figure 2** – a) Pulse scheme of the experiment with the relevant energy levels: The 4.7 eV pump pulse (cyan arrows) generates a coherent superposition of vibrational levels in the HOMO by impulsive stimulated Raman scattering (ISRS). These modes are then probed with element-selectivity by the soft X-ray pulse (blue arrows) tuned to the carbon K-edge. b) the experimental lay-out. A 786 nm femtosecond laser pulse is split and delivered to an optical parametric amplifier (OPA) where it is up converted to generate the ~4.76 eV pulse that initiate the Free Electron Laser process, as well as a frequency-tripling to generate the 4.7 eV pump pulse. The seeded electron bunch is propagated through a chain of apple II undulators, allowing control of the polarization and energy of the soft X-ray photons.

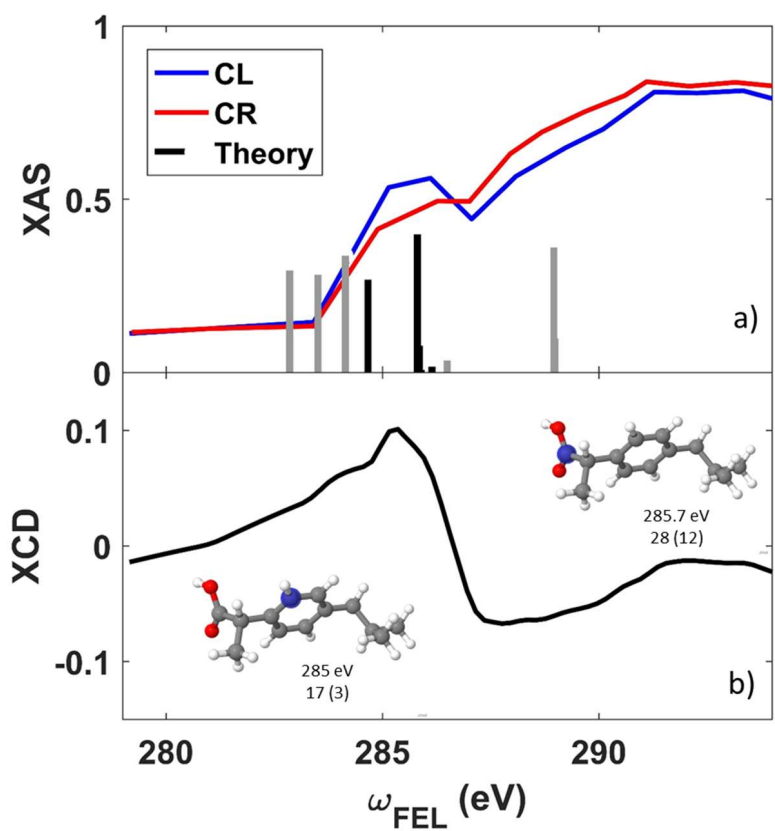

**Figure 3** – a) Steady-state Carbon K-edge absorption spectrum of S-IBP recorded with circularly left (blue trace) and right (red trace) polarized light. The sticks represent the calculated transition position and relative strengths. Black sticks are the transitions investigated in the present work. Grey sticks are other Carbons core transitions. b) Steady-state Carbon K-edge X-ray circular dichroism (XCD) signal calculated as the difference of the L and R spectra shown in panel a), normalized by their sum. Insets in panel b) show the selected atoms by the corresponding probe wavelength and their numbering. Numbers in brackets refer to the S enantiomer.

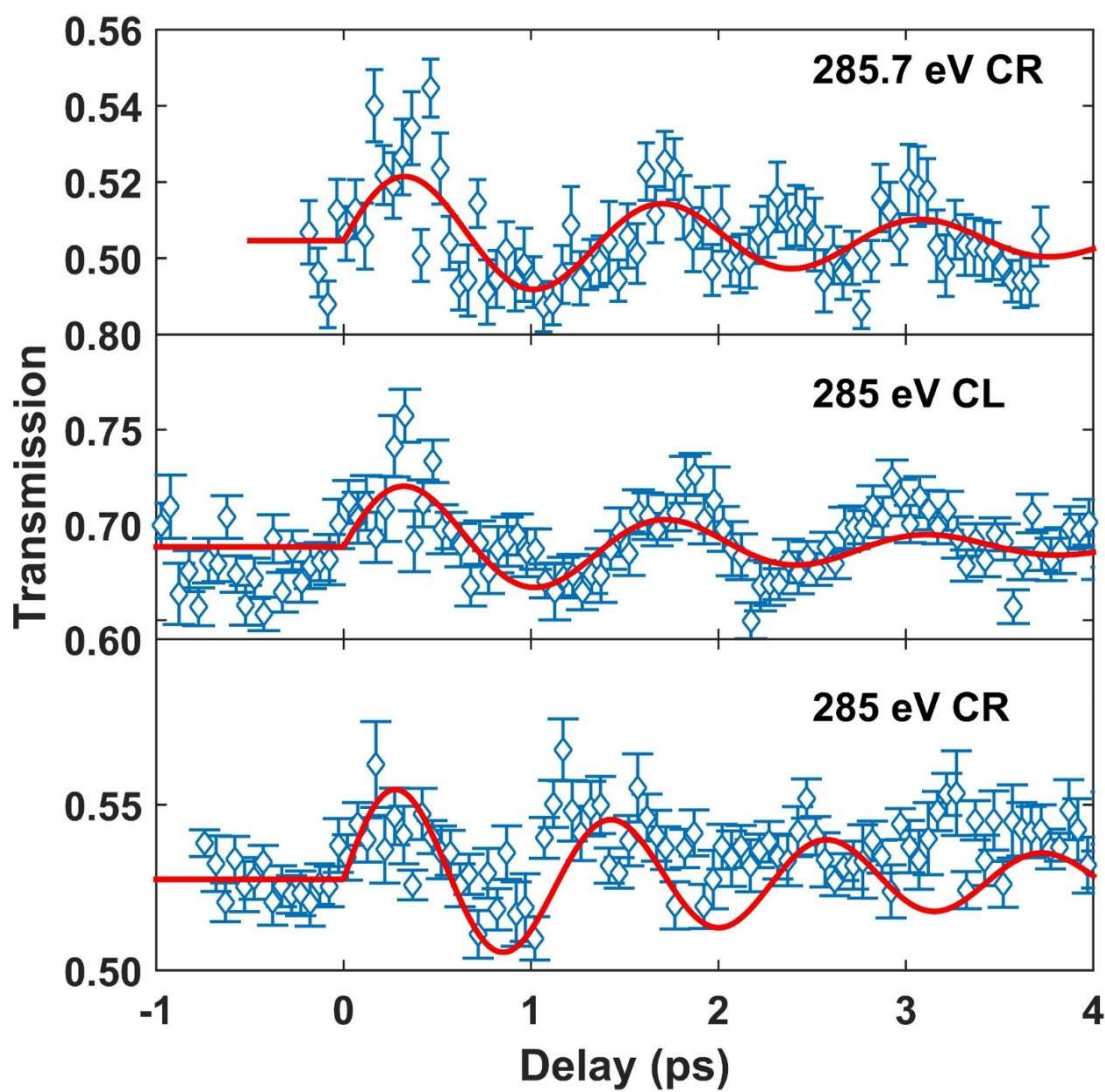

**Figure 4** – Time-resolved X-ray transmission signal of the racemic IBP sample measured using probe pulses of : a) 285.7 eV with circular right polarization: b) 285 eV with circular left polarization and; c) 285 eV with circular right polarization. Red lines are fits to the data with damped sinusoidal functions whose parameters are given in table 1 (See SI for further information).